\documentclass{iaus}
\usepackage{graphicx}

\title[Planetary nebulae and multiple evolutionary scenarios] 
{Are planetary nebulae derived from \\
multiple evolutionary scenarios?
}

\author[D.\,J.~Frew \& Q.\,A.~Parker]   
{D.\,J.~Frew$^{1,2}$ and Q.\,A.~Parker$^{1,2,3}$}

\affiliation{$^1$Department of Physics and Astronomy, Macquarie University, Sydney, NSW 2109, Australia \\[\affilskip]
$^2$Macquarie University Research Centre in Astronomy, Astrophysics \& Astrophotonics\\[\affilskip]
$^3$Australian Astronomical Observatory, PO Box 296, Epping, NSW
1710, Australia\\email: {\tt david.frew@mq.edu.au}}

\pubyear{2012}
\volume{283}  
\pagerange{119--126}
 \jname{Planetary Nebulae: an Eye to the Future} \editors{Arturo Manchado, Letizia Stanghellini \& Detlef Sch\"onberner, eds.}
\begin{document}

\maketitle

\begin{abstract}
Our understanding of planetary nebulae has been significantly enhanced as a result of several recent large surveys (Parker et al., these proceedings).  These new discoveries suggest that the `PN phenomenon' is in fact more heterogeneous than previously envisaged. Even after the careful elimination of mimics from Galactic PN catalogues, there remains a surprising diversity in the population of PNe and especially their central stars.  Indeed, several evolutionary scenarios are implicated in the formation of objects presently catalogued as PNe. We provide a summary of these evolutionary pathways and give examples of each. Eventually, a full census of local PNe can be used to confront both stellar evolution theory and population synthesis models. 
\end{abstract}

\firstsection  
\section{Introduction}
Planetary nebulae (PNe) are an important phase in the life of intermediate-mass stars.  While their formation is broadly understood, the mechanism(s) required to create the multitude of central star spectral types and PN morphologies remain unclear.   To help address these problems, we have compiled a volume-limited sample of PNe out to 3.0~kpc from the Sun, with the ambitious goal of measuring the fundamental properties of every PN and its central star (CSPN) within that volume.

Though not easy to produce, volume-limited samples are fundamental in astronomy, so a complete local PN census is needed to fully understand the Galactic disk population and the characteristics of their ionizing stars.   Surprisingly, there is a lack of consensus over the precise definition of a PN, a situation that lingers even after years of intensive effort
(Frew \& Parker 2010, 2011).  We argue that the `PN' term should be restricted to the ionized shell ejected after the AGB phase, or via a common-envelope (CE) event (De Marco 2009).  We make this point, since in many symbiotic systems, which can be confused with PNe, the ionized gas derives from a companion star, and not from the precursor of the white dwarf (WD) that is usually present in these systems, and whose own PN has faded.

\section{The Diversity of Central Stars}\label{sec:diversity}

There is a great deal of variation in the chemical compositions and effective temperatures, and hence spectral types, of nearby CSPNe. About 20\% are H-deficient, including those where Wolf-Rayet features are present.  Almost all belong to the [WC]--[WO] sequence, though recent discoveries are altering this paradigm.  The variable [WN4.5] nucleus of N~66, a peculiar PN in the LMC (Pe\~na et al. 1995), may be the product of an exotic binary evolution channel (Hamann et al. 2003; Lepo \& van Kerkwijk 2012). Certainly, its point-symmetric morphology (Fig. 1) is similar to known post-CE PNe (e.g. Miszalski et al. 2009).  Closer to home,  the faint nebula Abell~48 may host the first Galactic H-deficient [WN] star (DePew et al. 2011; Frew et al. 2012, in preparation), since PHR~1619-4913 (Morgan et al. 2003) is probably a ring nebula around the massive WN star PMR~5 (Todt et al. 2010a).   In addition, a [WN/WC] class has been proposed with PB~8  as the archetype (Todt et al. 2010a,b).  There are three more candidate [WN/WC] stars (Parker \& Morgan 2003; Parker et al., in prep.) that we are following up, but their origin is currently unclear (also see Werner 2012, these proceedings).

We also note the weak emission-line stars (`WELS';   Tylenda et al. 1993) which show narrow lines from the $\lambda$4650\,NIII-CIII-CIV blend and the CIV\,$\lambda$5806 doublet; CIII\,$\lambda$5696 is absent or weak, while HeII $\lambda$4686 emission is often seen.  These lines are also seen in some massive O-type stars, low-mass X-ray binaries and cataclysmic variables, so there is potential for misclassification.  The `WELS' appear to be a mixed group (Gesicki et al. 2006), including at least two different sub-classes: (i) H-rich stars with weak lines (some would be classed as Of or O(f) at sufficient resolution), and (ii) H-poor stars with weaker winds  than the [WR]s.    Since several CSPNe have been classified as both Of and `WELS' in the literature,  a correct classification may depend on the spectral coverage, resolution and sensitivity.  Even though the `WELS'  are heterogeneous, Fogel et al. (2003) and Girard et al. (2006) found that their PNe have lower mean N/O ratios, indicating they derive from less-massive progenitors, as suggested by their average $|z|$ height (see Table~1).  
Additionally, several close-binary (post-CE) CSPNe show emission from the $\lambda$4650 blend and CIV\,$\lambda$5806.  In these cases, the emission lines arise from the heated hemisphere of a cool companion to the true ionizing star (e.g. Corradi et al. 2011). 


\section{Planetary Nebulae as a Heterogeneous Class}\label{sec:hetero}  
The main point to take from the previous discussion is the wide diversity of emission-line CSPNe.  However, even armed with modern multi-wavelength diagnostic tools to weed out the mimics that have contaminated PN catalogues in the past (Frew \& Parker 2010; Frew et al. 2010), the observed diversity seen in PNe and their ionizing stars indicates the `PN phenomenon' is heterogeneous.  It is likely that several stellar evolutionary pathways form objects classified as PNe  (Frew \& Parker 2010, 2011), which include:\\
\noindent $\bullet$ \textit{Post-AGB evolution of a single star}, producing a classical PN.\\
\noindent $\bullet$ \textit{Short-period interacting binaries} ($P$ $\simeq$ 0.2 -- 20\,days) that have passed through a CE phase (De Marco et al. 2008).  Post-CE PNe appear to have quite distinct observational properties (Bond \& Livio 1990; Frew \& Parker 2007; Miszalski et al. 2009).  Moreover, some post-CE PNe may contain double degenerate cores (e.g. Danehkar et al. 2012), of interest because they are potential SN~Ia progenitors; there are even likely PNe around the classical novae GK~Per and V458~Vul (e.g. Wesson et al. 2008). \\
\noindent $\bullet$ \textit{Longer-period interacting binaries}, which may produce a range of phenomena, including circumstellar disks and collimated outflows.  A possible example is NGC~6302, whose rapidly expanding bipolar lobes were ejected 2000 years ago (Szyszka et al. 2011).  Soker \& Kashi (2012) speculate this is the product of a binary-induced \textit{intermediate luminosity optical transient}, an idea that holds promise.  A related object might be He~2-111 (Fig.~1).  There are also highly collimated nebulae which host B[e] nuclei (e.g. Mz~3), with or without symbiotic characteristics.  Lastly, the  yellow symbiotic stars (Frankowski \& Jorissen 2007) may be tentatively included here, plus the EGB~6-like CSPNe, associated with dense, ionized circumstellar nebulae (Frew \& Parker 2010; Miszalski et al. 2011).\\
\noindent $\bullet$ \textit{Pathways that form H-deficient nuclei}, such as the born-again scenario (Werner \& Herwig 2006), where a very late helium flash sends a WD star back to the AGB, forming a H-deficient CSPN.  It is possible that there are several alternative post-AGB channels leading to H-poor CSPNe (De Marco 2002), including binary-induced mixing (De Marco \& Soker 2002) and the diffusion-induced nova scenario (Miller Bertolami et al. 2011). \\ 
\noindent $\bullet$ \textit{Formation of PNe with O(He) nuclei}.  These possibly descend from the R~CrB stars, that may derive in turn from lower-mass WD mergers (Clayton et al. 2007). Alternatively, they are related to the [WN/WC] stars, since two O(He) stars have similar abundances to the CSPN of PB~8 (Werner 2012). Moreover, any suggested relationships between the R~CrB stars and the dusty [WCL] CSPNe are not yet clear (Clayton et al. 2011).\\
\noindent $\bullet$ \textit{Putative PNe/WDs from super-AGB stars}.   The SAGB stars derive from the heavier (6--8\,$M_{\odot}$) intermediate-mass stars, which undergo core C burning to form O-Ne WDs; H1504+65 might be an example (Althaus et al. 2009; Werner 2012), though it has no surrounding PN, while the `hot DQ' stars (Dufour et al. 2008), and two cool O-rich WDs found by G\"ansicke et al. (2010), are possible evolved O-Ne cores and also of great interest.

\begin{figure}[tb]
\begin{center}
\includegraphics[width=12.45cm]{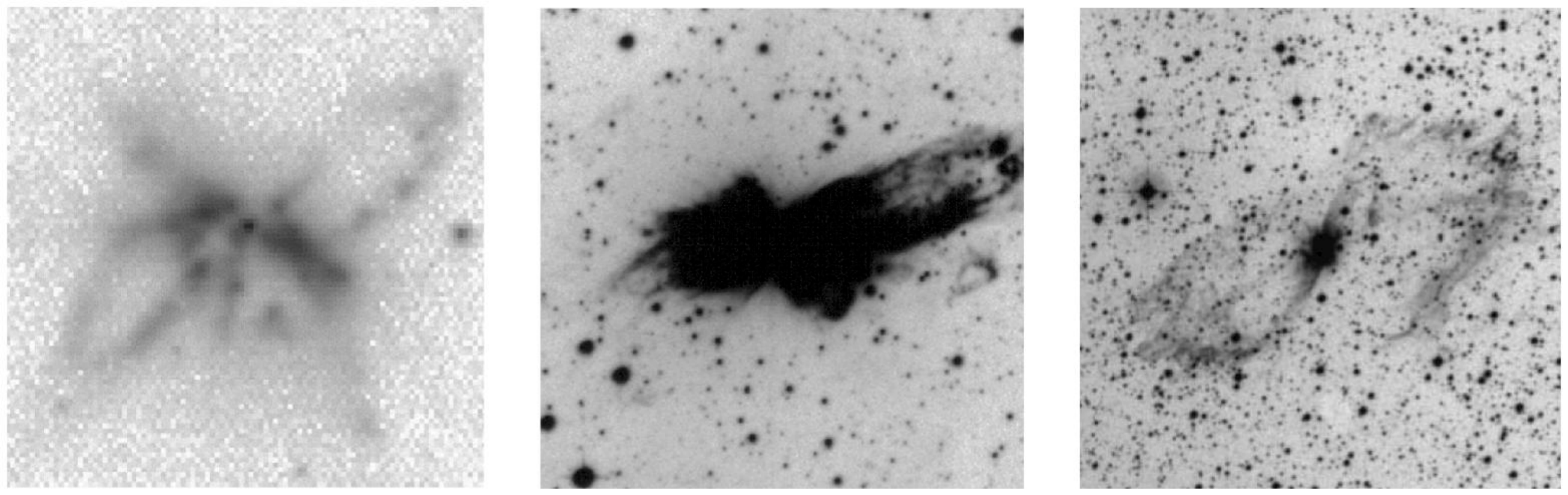}
\end{center}
\caption{Three unusual PNe with collimated outflows. The left panel shows a narrowband $[$O\textsc{iii}$]$ image of LMC~N66 taken from Shaw et al. (2006).  The middle and right panels are narrowband H$\alpha$~+~$[$N\textsc{ii}$]$ images (Parker et al. 2005) of NGC~6302 and He~2-111 respectively. } \label{fig_morph}
\end{figure}

\begin{table}[tbp]
\begin{center}
\caption{Preliminary mean $|z|$ distances for various subsets of Galactic disk CSPNe.}
\label{table:z_heights}
{\scriptsize
\begin{tabular}{llc|llc}
\hline
Class                               &     ~~$|z|$ (pc)     & $n$~ & ~~~Class    & ~~$|z|$ (pc)   &    $n$ \\
\hline			
O(H), DAO, DA~~     & 300 $\pm$ 60 &105~~~&~~~PG\,1159, PG\,1159h~~~~   & 380 $\pm$ 80  & 20  \\
$[$WCL]                               & 160 $\pm$ 50  & 9~~  &~~~`WELS'  &	 380 $\pm$ 80   &  21 \\		
$[$WCE], [WO]                  & 200 $\pm$ 40 & 19~~ & ~~~O(He)   & 1200 $\pm$ 400~~~  & 2  \\
$[$WNC]                        & 250 $\pm$ 100~~   & 4~~  & ~~~B[e], symbiotics   & 160 $\pm$ 50    & 7  \\
$[$WC]-PG\,1159~~     &  750 $\pm$ 300  &   2~~   &  ~~~post-CE 	  &	 230 $\pm$ 50  &	19   \\   
\hline
\end{tabular}
}
\end{center}
\end{table}

\section{Progenitor Populations}
Volume-limited samples allow us to investigate the mean $|z|$ heights, and hence progenitor ages and masses of different CSPN populations for the first time; see Frew \& Parker (2011)  for approximate frequencies of the different types.  Table~\ref{table:z_heights} summarises some preliminary results for a 3-kpc volume-limited disk sample ($n$ = 425\,PNe), which has a mean  $|z|$  height of 240\,$\pm$\,40\,pc.  We find some interesting trends, with the caveat that only $\sim$50\% of nearby CSPNe have spectra, so there is sample bias; e.g. the PG1159s are less luminous than the [WC]s, so are less likely to be observed if close to the Galactic plane.   While several groups suffer from small number statistics, it is probable that the two born-again [WC]-PG1159 stars in A30 and A78 come from lower-mass progenitors. Similarly, the O(He) stars may derive from the R~CrB stars rather than the [WC/WN] path, looking at the scanty available data.  We also caution that besides the `WELS', other classes are heterogeneous, such as the B[e] stars, PG\,1159s and probably the [WN/WC]s.   Obviously, larger samples are needed before more definite conclusions can be made.

\section{Future Work}
Galactic PNe currently number nearly 3000 (Frew \& Parker 2010; Parker et al., these proceedings), with more discoveries expected in the near future.   The role of binarity in PN formation is becoming clearer, so continued surveys for new binaries (e.g. Douchin et al., these proceedings) are needed to refine the post-CE frequency in the Galactic disk.   A detailed spectroscopic survey of all local CSPNe is also necessary, combined with multiwavelength observations of unusual nuclei in order to confront evolutionary theory.   Especially interesting are the `WELS', as new examples of [WNC] and [WN] stars may be lurking there.  We look forward to the day when a precise taxonomy of PNe and their ionizing stars is attained, but much more work is still to be done.

\end{document}